# A review of Smart Contract Blockchain Based on Multi-Criteria Analysis: Challenges and Motivations

**Norah M. Alshahrani[1, 2, *], M.L. Mat Kiah[1, *], B. B. Zaidan[3], A. H. Alamoodi[4] and Abdu Saif [5]**

[1]Department of Computer System & Technology, Faculty of Computer Science & IT, University of Malaya, 50603 Lembah Pantai, Kuala Lumpur, Malaysia
[2]Department of Computer Science, Applied Collage, King Khalid University, P.O. Box 62569, Abha, Saudi Arabia
[3]Future Technology Research Center, National Yunlin University of Science and Technology, 123 University Road, Section 3, 64002 Douliou, Yunlin, Taiwan
[4]Department of Computing, Faculty of Arts, Computing and Creative Industry (FSKIK), Universiti Pendidikan Sultan Idris, Tanjung Malim, Malaysia
[5]Department of Electrical Engineering, Faculty of Engineering, University of Malaya, Kuala Lumpur, 50603, Malaysia

[*]Corresponding Author: Norah M. Alshahrani. Email: nshahrani@kku.edu.sa
M.L. Mat Kiah.Emails: misslaiha@um.edu.my

**Abstract:** A smart contract is a digital program of transaction protocol (rules of contract) based on the consensus architecture of blockchain. Smart contracts with Blockchain are modern technologies that have gained enormous attention in scientific and practical applications. A smart contract is the central aspect of a blockchain that facilitates blockchain as a platform outside the cryptocurrency spectrum. The development of blockchain technology, with a focus on smart contracts, has advanced significantly in recent years. However, research on the smart contract idea has weaknesses in the implementation sectors based on a decentralized network that shares an identical state. This paper extensively reviews smart contracts based on multi-criteria analysis, challenges, and motivations. Therefore, implementing blockchain in multi-criteria research is required to increase the efficiency of interaction between users via supporting information exchange with high trust. Implementing blockchain in the multi-criteria analysis is necessary to increase the efficiency of interaction between users via supporting information exchange and with high confidence, detecting malfunctioning, helping users with performance issues, reaching a consensus, deploying distributed solutions and allocating plans, tasks and joint missions. The smart contract with decision-making performance, planning and execution improves the implementation based on efficiency, sustainability and management. Furthermore, the uncertainty and supply chain performance lead to improved users' confidence in offering new solutions in exchange for problems in smart contacts. Evaluation includes code analysis and performance, while development performance can be under development.

# 1  Introduction

Blockchain technology has gradually gained popularity due to the blooming interest in implementing the Internet of Things (IoT) in various fields of application, such as smart devices and artificial intelligence (AI). Intensive research and implementation of blockchain technology resulted from new ideas for applications such as smart logistic management, smart cities, smart contract and more to be listed. The significant advantages of employing blockchain technology are that the information cannot be erased, modified and transparent while eliminating the necessity of a third party or central authority [1]. As a result, the cost and duration required for the transaction reduce drastically. A smart contract is executed in a computer protocol that utilizes blockchain technology as a Virtual Machine (VM) [2]. Even though the smart contract can be used by multiple sectors, such as financial institutes and engineering, implementing the applications is challenging. The field of application for a smart contract will determine the type of VM applied, such as Ethereum and Corda [2]. Each VM must have different programming languages and blockchain networks to operate and achieve the implementation goals. Hence, by employing a suitable VM, the performance of executing a smart contract can drastically increase.

Therefore, smart contracts have essential criteria that must be determined initially. However, the emphasized critical criteria vary for different fields of application. Then, Multi-Criterion Decision Method (MCDM), an evaluation approach, is used to streamline the evaluation process. There are various technologies for a smart contract platform like Bitcoin, Ethereum, Counterparty, Stellar, Monax and Lisk [3] and the superior platform is limited over another's. In this context, a smart contract platform selection is a complex multi-criteria problem due to the multiple criteria and criteria conflict. Several criteria to evaluate smart contracts include privacy, security, performance, .etc. The decision-making process gets more complicated as the number of decision alternatives and criteria increases [4]. There is a unified evaluation matrix for smart contract platforms' evaluation and selection problems.

In addition, the preferences (criteria weight) is yet set for the unified evaluation matrix. MCDM helps the decision maker to obtain the best solution based on the critical criteria the decision maker emphasized in a set of conflicting critical criteria[5]. Furthermore, by adopting the MCDM method, decision-makers can be more accurate and faster, increasing the overall performance of a process. As a result, implementing a project's new idea or transitioning process can be hastened. Numerous MCDM methods are available such as Multi-Attribute Utility Theory (MAUT), Analytic Hierarchy Process (AHP) and Fuzzy Theory. Nevertheless, every MCDM method inherits various advantages, weaknesses and implementation sectors. Thus, selecting a suitable MCDM method for utilization is essential.

## 1.1 Internet of Things (IoT)

Internet of things (IoT) is the concept of connecting any network-enabled devices, known as "smart devices" or "smart appliances," to the internet and other connected devices at any time and any place [6]. Mobile phones, digital tablets, refrigerators and home security systems are examples of smart devices and smart appliances. The smart device can "sense" its environment, process information and share or exchange the data gathered with other devices through network connection. Smart devices are connected to sensors or actuators for the smart devices to "sense" the environment. In contrast, to broadcast information or data, the smart device is connected to the internet through a Wi-Fi connection, Bluetooth connection, infrared connection, or Near-Field Communication (NFC). The blockchain in the IoT acts for reliable for sensitive information of smart cities networks [7].

Hence, the concept of IoT and blockchain was developed and being applied widely in different areas of industries and applications, including finance, healthcare, utilities, real estate, logistics, education, government sectors, smart contract, smart cities and cryptocurrency [8], [9],[10].

*1.2 Blockchain*

   Blockchain is a peer-to-peer public distributed ledger software that records transactions, agreements, contracts and sales while verifying all the records by the majority or public [11],[12]. A blockchain is a growing list (known as a chain) of records or transactions (known as the block) which are linked together from one block to another in sequential order using a cryptography algorithm [13],[14]. Due to the information being stored in a cryptography method, the transaction is secure within the blockchain [15]. Then, before a transaction or block is linked to the blockchain, it is verified by a majority in the distributed computing system via a consensus mechanism [16],[17]. This verification process evidences the occurrence of the transaction and prevents the information from being erased, contaminated, or modified. Thus, the information block within the blockchain is immutable. Each verified block contains a hash value of the previous block[18],[19]. Therefore, any new block reference to the last block creates a chain back to the first or the root block, forming an information grid [20], as shown in Fig. 1. This network of information grids displays the data's origin and destination, enabling transparent authentication and inspection of all transactions. In addition, Fig.1 illustrates the advantages of blockchain technology for records transactions and agreements, in smart contracts issues to the data's origin and destination. There are several ways to find the data's origin and destination using Block Data and Block Header. In these cases, blockchain is used to turn alert, share information, decentralization, update the Block Data and Block Header, to share location during network platform. Consensus algorithms and sharing techniques are required to improve blockchain performance in transaction speed, scalability, network partitioning and malfunctioning takes.

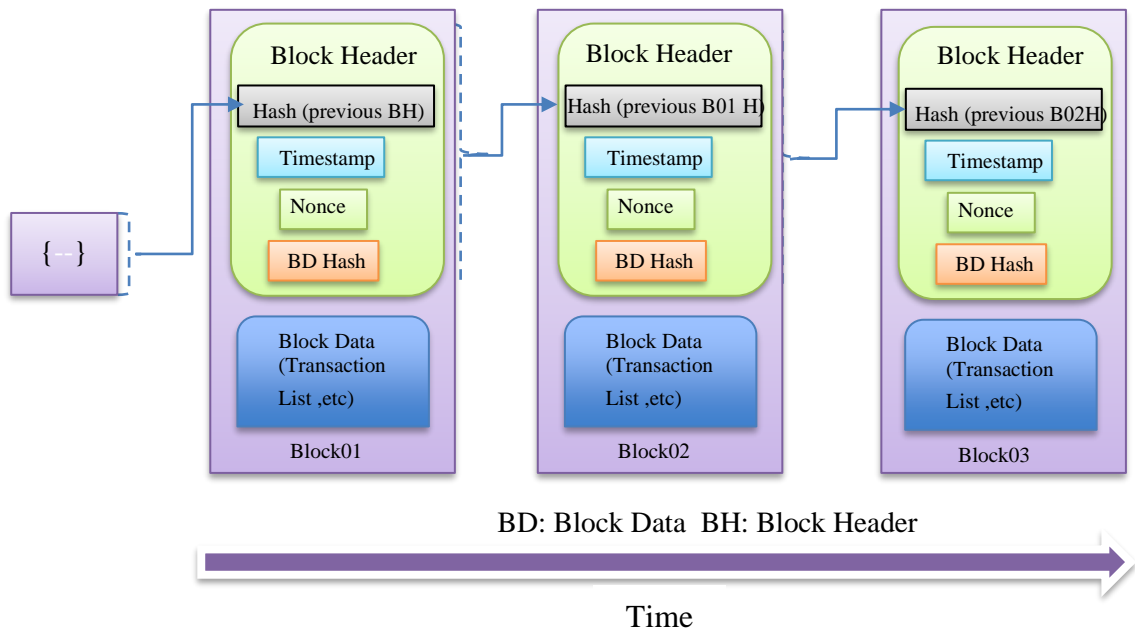

**Figure 1**: Block Schematic Chain

With the verification of the majority within the blockchain information grid, the blockchain can operate without the necessity of a trusted third party or central authority [21]. The majority verification is the key advantage of blockchain, also known as decentralized used an application interface with a shared central database[22]. Blockchain technology's ultimate goal is this decentralized environment where there is no involvement of a third party within the transaction[23]. As a result, all transactions can be completed at a lesser cost, with faster transaction speed and a safer environment. Moreover, all historical transactions are auditable and traceable[24]. In addition, doubt about information integrity can be eliminated from the reliable system created by the advantages of blockchain. This system is also suitable for implementing smart contracts.

*1.3 Smart Contract*

A smart contract is a digital program of transaction protocol (rules of contract) based on the consensus architecture of blockchain. The protocol is deployed in the blockchain [25] and can self-execute as the agreement is met. A smart contract can automatically perform calculations, storage information, transfer, etc. Furthermore, it can support polymorphism and inheritance [20]. Data authorization rules, functions and processes are embedded in smart contracts. Therefore, decentralized can be managed via smart contracts, reducing the cost processes significantly. Evaluation and development are types of smart contracts. Evaluation includes code analysis and performance, while development performance can be under development. Ethereum addressed Solidity [26] as a language (including code instructions and event state (data) like the initial, intermediate and also final) to implement smart contracts.

Furthermore, the authors of [27] introduced smart contracts as part of the Ethereum virtual machine. Ethereum virtual machine is an isolated environment while accessing data among smart contracts is limited. The transactions contain outputs of smart contract code. Transactions code execute within the Ethereum virtual machine. A smart contract can perform actions, including data collection, processing and adopting specific solutions. A smart contract is based on blockchain; the arrangement is emulated in the programming language. Then, the smart contract transfers to a blockchain, which will be automatically self-executed as the agreement are reached. Party is not able to globally prevent the smart contract execution. A smart contract is an automatic program placed in a specific address in the blockchain to perform processes accordingly. Once the particular event occurs and the transaction reaches the smart contract, the blockchain-distributed virtual machine executes the program's code. A new party can join a smart contract and initiate automatic execution by meeting specific historical conditions[28].

The development of blockchain technology enables smart contracts in a different field, which it initially proposed by Nick Szabo in the 1990s[29]. Blockchain networks act as VM to execute smart contracts[30]. A smart contract is "a computerized transaction protocol that executes the term of a contract"[31]. To be exact, a smart contract is a digitalized way of executing contracts where two anonymous individuals digitally sign and verify the smart contract in the blockchain network. The smart contract's contract provisions are expressed in a computer programme and execution happens automatically when specific criteria are satisfied. After execution, the smart contract transactions are stored, replicated and updated in the distributed blockchain information network [32]. As a smart contract can self-execute, the intervention of trusted intermediaries between the transacting parties can be eliminated[33]. However, smart contract modification after execution is prohibited [34]. With new programming tools, developing a smart contract for implementation reduces complexity. Reducing administration and service costs while increasing the efficiency of a business process are the advantages of adopting smart contracts[28].

Consequently, the occurrence of financial fraud will be significantly mitigated. Two types of smart contracts were identified: strong and weak smart contracts [34]. A strong, smart contract is temper proof by the legal court and has a prohibitive cost of revocation and modification[35].

In contrast, the weak smart contract does not. A smart contract applies to both physical and non-physical assets such as financial [36], car, land and property [37] , cryptocurrency application [38], legal commitment [39],[8] and voting system[40]. A smart contract can operate in different blockchain networks, such as Ethereum [41], Corda, Hyperledger Fabric, Stellar, Roostock and Electro-Optical System (EOS) [42], with different selection criteria.

Therefore, distributed decision-making algorithms (DDM) have been applied in the development of smart contract blockchain applications such as collection mapping [43] and dynamic task allocation [44], avoiding obstacles. However, DDM algorithms are an open problem in deploying massive numbers of smart contracts [45]. Furthermore, flexible, and autonomous multi-criteria analysis for decision-making using DDM algorithms is needed to tackle the industry's new challenges. However, blockchain ensures that all smart contract in a decentralized network shares an identical state. For example, blockchain creates distributed voting among multi-criteria analyses requiring an agreement.

Moreover, blockchain technology achieves collaboration models between heterogenous multi-criteria analysis. Further, adopting blockchain technology in distributed decision-making can allow operators and maintainers to collaborate on smart contracts. Furthermore, blockchain is used widely over AI due to its unique advantages in surveillance [46]. Blockchain advantages include decentralized, immutable, deterministic and data integrity [47]. Because all transactions and agreements are stored in blockchain, there is no requirement to invest time in training and learning phases in joining a new multi-criteria analysis collaboration [48]. Then, further smart contact can synchronize automatically with multi-criteria analysis via downloading all previously stored historical events from the blockchain.

*1.4 Smart Contract Evaluation*

The smart contract implementation involves complex criteria selection processes such as the implementation blockchain network and the critical criteria of the smart contract. Each blockchain network has its advantages over a specific field of application. For example, corda specializes in digital currency applications, Hyperledger Fabric supports available enterprise applications, while Ethereum applies to various applications[49]. Thus, the blockchain network implementation should be determined by the nature application of the smart contract[32]. Therefore, the intention for the application of the smart contract is crucial[50]. Moreover, blockchain network criteria such as execution environment, supported languages, turing completeness, data model, consensus algorithms and permission depend on the selected blockchain network[51], as shown in Fig. 1.

The critical criteria that hinder smart contract implementation are security, consensus protocol, transaction speed, scalability, versioning, re-entrancy, cost and privacy [38],[52]. The weightage for every key criterion will change depending on the smart contract application. For example, the security criteria of a smart contract will be important for sensitive and confidential information. In contrast, transaction made within a shop emphasizes the transaction speed criterion. However, in most smart contracts, the primary concern criteria are security, transaction speed and consensus protocol.

Security is a significant concern not only for the smart contract but also for the entire blockchain network. Due to the immutable advantage of blockchain, smart contract transaction within the blockchain network is secure. The decentralized authentication rules and logic of blockchain increase the authentication efficiency of smart contracts compared to traditional authorization protocols such as Role Based Access Management (RBAC), OAuth 2.0, OpenID and LWM2M[53],[54].

Transaction speed is the duration of a blockchain network completing one transaction. The duration to complete one transaction is estimated to be ten minutes[55]. The transaction speed is proportional to the security efficiency of the blockchain network. The higher efficiency of the security, the duration required to complete one transaction increases[56]. Moreover, the transaction speed escalated with the increased complexity of the smart contract [57]. The consensus protocol is a mechanism where all miners agree with the same message within the blockchain network to ensure the latest block is correctly added to the chain [58]. Moreover, consensus protocol protects blockchain networks from malicious attacks. Proof of work (PoW) and Proof of Stake (PoS) are part of the consensus protocol for blockchain[59]. A valid PoW is generated when the block header hash value is less than a set value. This process consumes a tremendous amount of electricity and computational power[60]. In contrast, PoS compares the resources based on the percentage of cryptocurrency held by the miner [61]. PoS provides additional protection from malicious attacks by increasing the attack cost because the attacker requires a near majority of cryptocurrency to initiate the attack[62].

*1.5 Multi-Criteria Analysis*

The MCDM method is used to find the optimum solution depending on the criteria required by the decision-maker. MCDM is defined as evaluating multiple conflicting criteria while considering the decision criteria needed to determine the most efficient decision[63],[64]. Multi-Attribute Utility Theory (MAUT), Analytic Hierarchy Process (AHP), Fuzzy Set Theory, Case-based Reasoning (CBR), Data Envelopment Analysis (DEA), Simple-Attribute Rating Technique, Goal Programming, ELECTRE, PROMETHEE, Simple Additive Weighting (SAW) and Technique for Order of Preference by similarity to Ideal Solution (TOPSIS) were the eleven types of MCDM methods identified [65],[66]. Each of the specified MCDM methods cannot apply to every application. For example, the suitable MCDM methods in the engineering field are CBR, TOPSIS and Fuzzy Theory due to the limited information available and the benchmarking process to the ideal cases [67], [68],[69]. In addition, MCDM evolution based setting assessment criteria, weights, which compared with the conventional cognitive learning process [70].

However, applying the Analytical Hierarchical Process (AHP) method is appropriate for the public policy-producing sector because it is scalable[71]. Even though the MCDM methods are suitable to use in the specific field, they do inherit some setbacks, such as required drastic input, sensitivity to inconsistent data, etc. Thus, the MCDM method's implementation depends on the field of application.

*1.6 Motivation*

In a multi-criteria analysis network, centralized control suffers from a single point of failure, whereas decentralized control suffers from lacking global knowledge. Therefore, decision-making in the centralization robots network takes a long time to control blockchain smart contracts during task performance and the collaboration between users due to delays in response. Decentralized improves the performance of the smart contact and reduces the time spent on doing tasks. Furthermore, sharing information is essential to support the interaction of multi-user collaboration for operation in environmental exchange, uncertain conditions and external disturbances. A successful solution for multi-user interaction issues together to perform tasks and record event history by blockchain can improve the efficiency of multi-user interaction. Therefore, implementing blockchain in the multi-criteria analysis is required to increase the efficiency of interaction between users via supporting information exchange with high trust, detecting malfunctioning, helping users for detecting performance issues, reaching a consensus, deploying distributed solutions, allocating plans tasks and joint missions.

*1.7 Contribution and Scope*

A Multi-Criteria Analysis is used to make the research on blockchain for improving Smart contracts a very relevant and strategic topic. To combat a multi-Criteria Analysis efficiently and effectively, managing homogenous and heterogeneous smart contracts is necessary to avoid any mistakes, reduce response time, speed transactions and identify business scope in the cases earlier. Furthermore, due to the increasing number of Blockchain for combating smart contracts in different ways, multi-criteria analyses suffer from global information, malfunction, controlling collaboration and so on.

This conceptual research proposes focusing on avoiding network partitioning and improving the scalability of Smart Contract Blockchain performance, which can contribute clearly and more effectively to combat multi-Criteria Analysis tasks for managing a business. As the previous research on the same topic differs, this research mainly focuses on blockchain for managing a Smart Contract from the respective Challenges and Motivations. Blockchain makes data public for a network platform, enables information share and delivers data of user's relation to all in the same network. We focus mainly on discussing challenges in Implementation, Decision making, Policy and applications. In this context, the challenges focus on how to develop a decentralized ledger platform with corresponding algorithms to enable a multi-criteria analysis with a tolerance of network partition for Smart Contract Blockchain as follows:

*1.7.1 Multi-Criteria Analysis Management and Controlling*

We discuss the proposed Blockchain technology as a critical solution framework for managing and controlling smart contracts. Blockchain is proposed to control homogenous and heterogeneous networks during operation in uncertain and inappropriate environments. Each Blockchain user acts to share information with others via a smart contract. Then, the multi-criteria analysis behaviour management can be changed accordingly until a specific task is controlled effectively and efficiently. In this case, Blockchain is decentralized, which can handle many transactions produced from capable of applying in every application. Therefore, it notes that the interaction among multi-criteria analysis management can increase the efficiency of the smart contract.

*1.7.2 Decentralized of Multi-Criteria Analysis*

Blockchain is a decentralization network in which smart contact suffers from multi-criteria analysis. Moreover, blockchain helps the smart contract process transactions and store the world state equally. In the case of a multi-criteria analysis can perform their task efficiently. Furthermore, joining a new user to a group is more accessible via copying the smart contract and starting sharing events in the blockchain. Due to the blockchain's decentralization feature, joining new users and malfunctioning in the task cannot affect the performance of Decentralized of Multi-Criteria Analysis. Here, we discuss the characteristics of realistic Decentralized of Multi-Criteria Analysis for combating and how to develop the consensus algorithms considering consistency based on the current status. Further, we discuss the consensus algorithms with dynamic sharing technique to increase blockchain scalability so that the transaction processing and world state is limited in the sharing range.

**2 Structure**

The remainder of this paper is categorized as follows: introduction and background of smart contracts and blockchain in the new technologies. It follows, presents an overview of the challenges outlined and highlights the research motivation. Finally, the conclusion of the paper.

**3 Challenges**

This work employs a review to ensure accurate and impartial data search and retrieval. This study summarises pre-existing research that discusses various challenges realized by researchers. The difficulties identified considered implementation, decision-making, policy and application. All these challenges are addressed in Fig. 1. as follows.

*3.1 Implementation*

This section aims to address and discuss some of the main aspects of this study. Many challenges jeopardize various research efforts in scientific fields and addressing them is warranted in this area of research. The section mainly addresses the implementation of blockchain in the context of [smart contract]. Various challenges are connected with implementation, so we need to discourse them. Hence, the system updates concerning the system's architecture are costly [72]. It is clear that such a challenge comes with its sub-challenges (i.e., issues) and these issues discuss (1) System Implementation, (2) Security implementation and (3) Integrity of Implementation. All these sub-implementation challenges are discussed below.

*3.1.1 System Implementation*

The issues in system implementation. Authors have stressed various pitfalls and among their main concerns was the difficulty in understanding the nature of some complex blockchain systems [56]. Other authors have also discussed Supply Chain Management (SCM) request combinations at several levels and nodes, which makes implementing blockchain systems a heavy process [73]; another system implementation challenge was the management and scheduling of this system after implementation [74, 75]. The system implementation issues are considered from different aspects to make the blockchain more viable for general purposes[76]. Another critical challenge was discussing the system implementation need for unified connectivity, which might be difficult to maintain and track, especially in systems such as blockchain to vehicle charging units [77]. Moreover, system implementation can be challenging because of the structure of the organizations' layers [78] and the variety of structures, which can be challenging, especially in distributed locations [79]. The last issues in this category were attributed to users' need for a high level of transparency, let alone having a centralized authentication system, which can be challenging to implement, especially in large systems [80].

*3.1.2 Security Implementation*

The security implementation challenges are used in evaluating the blockchain's performance in the context of smart contracts. Previous researchers have raised various considerations based on this review. The security implementation issues in that context focused mainly on different layers in the structure where every level requires different security systems [81]. Other authors have focused on various services with different security and reliability requirements [82]. More implementation security concerns were attributed to system architecture [83],[84]. The last group of security implementation challenges addresses aspects such as the security of information[85], the existence of loopholes at different levels for information tapering [86] and the weakness of current systems' security that prevents their effective utilization [87]. The last issue is attending security of information protection for the assets [88].

*3.1.3 Integrity of Implementation*

Based on the review work, this sub-section discusses the last class of challenges associated with implementation. Several issues have been identified. First, performance may be challenging because data integrity is at stake [86]. Furthermore, implementation integrity in the blockchain domain faces several other risks, including selecting algorithms that should consider various issues [89]. Ensuring stakeholders' knowledge integrity is also worth considering [90], as nodes exchange and integrate data simultaneously [91].

*3.2 Decision Making*

Another important class of challenges is linked to decisions associated with blockchain technology. An analysis of the literature indicated that much as this area is significant, deciding to use or integrate it into daily life and aspects is still in its infancy. Therefore, the authors in this subsection illustrated their primary concern in decision-making. The planning and execution stage includes some of the factors in the decision.

*3.2.1 Planning*

For the first class of planning challenges, the main issues are circled factors that occur during the planning stage of any project, especially the blockchain ones. The challenges in blockchain projects for uncertainty about the nature or aspect of the project, including the possibility of having network instability, are considered significant issues [92]. In addition, a high level of uncertainty regarding digital technologies, especially integrating the blockchain with new development technologies [93]. On the other side of planning, issues involved uncertainty concerning some processes. Others were quite challenged by the existence of multiple uncertain factors that can change depending on the area of operation, making it difficult to generalize a framework [94].

Furthermore, uncertainty was presented in constructing a decision support model algorithm for blockchain that was rendered a complex task [95]. The last issue reported in the planning stage focused on factors other than uncertainty and was more connected to various features. Authors in this context discussed that having many variabilities to be considered in the decision-making of many functions, like functional language, can make the design and planning stage challenging [96]. Furthermore, decision selection challenges regarding individuals from various backgrounds and expertise were also an issue [97].

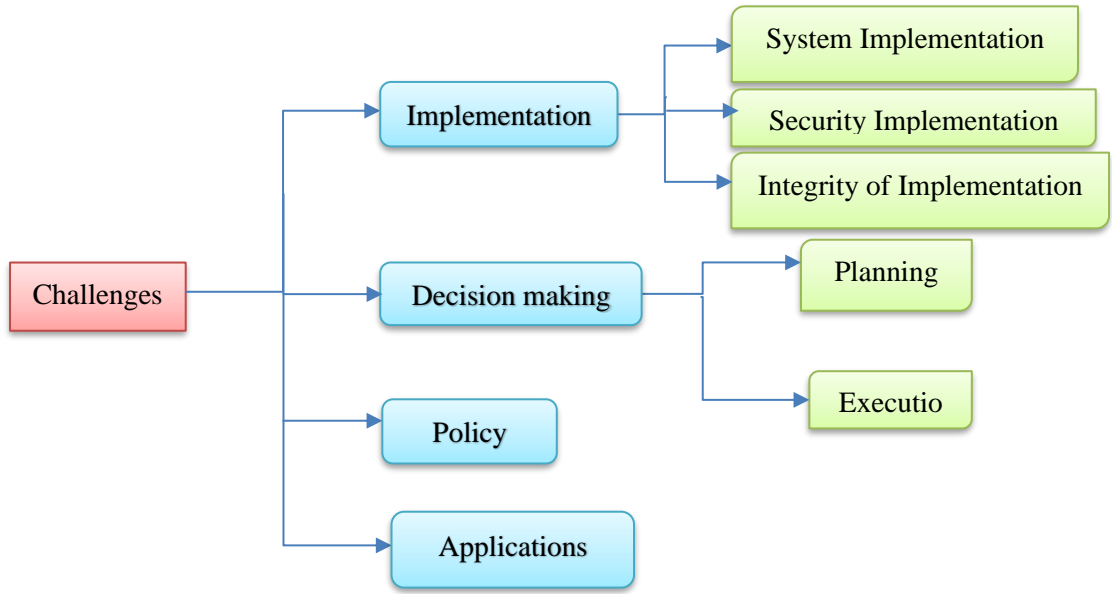

**Figure 2:** Classification of Challenges

*3.2.2 Execution*

Decision-making is not always about planning; it is also an issue that can be faced before and while deciding, especially when unexpected events arise. In the context of blockchain, execution can be faced with specific problems that require fast and good decision-making towards making the project survive and the issue it might face. The first issue reported in this sub-category discussed the system complexity during execution, especially in cases where two decision-making techniques may increase complexity and decrease effectiveness [98]. Therefore, during the decision-making, the main issue is selecting a platform with a specific context, different scenarios and stakeholders [99]. In this context, a platform includes many factors ranging from political and social to financial, which can complicate the decision-making process [100]. Other issues focused on the flexibility and versatility of digital manufacturing make it difficult to account for all determining factors in decision-making [101, 102].

The following set of problems associated with project execution is that during certain aspects where a decision is needed, it will be difficult when the decision-making is required in real-time [103]. Another challenge is globalization, which makes tracking and centralized decision-making much more complex [104]. The last reported issue concerns the revision of ranking decisions [105].

*3.3 Policy*

Another class of challenges was more concerned with ethics over a technicality, discussing issues that might touch on policy challenges with using blockchain technologies. Researchers in that capacity addressed policy issues concerning various factors, including preserving privacy [92] and the versatility of value chains considering the imposed policy framework and guidelines from national authorities [106]. Policy challenges were also discussed from a cultural perspective based on cross-cultural operations analysis [39]. Blockchain information systems in multinational organizations require stakeholders with various aims and scopes for different projects [107]. The variations of standards and requirements by customers based on their region [108], mainly when these variations of conditions occur in unusual cases like between dynamic environment and non-dynamic environment [105], [109], are also considered. Other policy challenges involved applying blockchain policies and studies cited the high requirement and need for precision and accuracy of the policy [100].

The need for organization and policy rules to be electronically processable while at the same time humanly understandable [110]. In addition, from a policy perspective, there should be constant cybersecurity vulnerability assessment frameworks [111] to reach a cybersecurity maturity while considering the existence of different requirements in the context of various applications [112] and the requirements of other blockchain transactions, such as digital contract [74]. The last set of issues concerning policies discussed aspects such as the trust factor that requires constant assessment when dealing with or implementing blockchain in a specific project [79] or when the critical decision of the technology is based on set protocols [90].

Moreover, the importance of a policy framework for future enhancement was also emphasised [113]. Finally, the policy requirement for technology where different factors are required to be considered across other times was also discussed [114]. Another significant challenge is the lack of government regulations [115]. Additionally , various applications utilizing the technology in the blockchain domain were apparent across the scientific literature, indicating how deep and integral blockchain is implemented. However, blockchain still faces the central issue in every type of technology, such as the issues and challenges hindering its ultimate utilization. Previous research has demonstrated the diversity of blockchain applications and the complexity attributed to the variations of techniques [82], [116]. Moreover, blockchain plays a vital role in applications [112, 117]. However, the generalization of blockchain is impossible and limits its applications and adherence to various standards [107].

**4 Motivations**

The motivation aims to identify some of the main aspects of this study. Various explanations in various fields, such as performance, improvements, uncertainty and supply chain, are connected to the current research and thus need to be discussed in Fig. 1. bellow:

*4.1 Performance*

The performance of the motivations derived from relevant research stimulated researchers in this field. Performance motivation can be divided into efficiency, sustainability and management. These performance motivations are discussed below.

*4.1.1 Efficiency*

Performance efficiency is considered the most notable benefit, especially in different and various examples of integrating blockchain technology systems. These prominent examples include monitoring and billing systems [74]. The other consideration of the motivations of blockchain efficiency is based on processes and a predictive algorithm [103]. It can also be observed in different strategies to increase efficiency and reduce cost, especially in digitization [93]. Another critical motivation was discussing the required transaction size for efficiency effects, which can benefit data protection and integrity [89]. Moreover, method analysis can be considered motivational because of its efficiency and profitability. Others were motivated by efficient performance from clustering techniques and service strategies [86].

The authors also discussed the importance of improving supply chain performance and efficiency [64]. Thus, many factors were considered to improve performance and influence energy efficiency in shipping operations [94]. The last group of efficiency performance motivations involves addressing aspects such as reliability, availability and throughput in areas integrated alongside blockchains like cloud computing [82] and the efficacy of machine learning algorithms [116]. Data interoperability effectiveness in data exchange has also been significant [118].

### 4.1.2 Sustainability

For performance sustainability, the core characteristic is to motivate the circular economy to involve sustainability and the present status of SCM virtually depends on sustainability motivation [73]. In intermodal transport networks, the management employs sustainability to improve the logistics and transport system. In blockchain products, decision-making improves through sustainability [95]. Technological advancement, such as cloud computing and big data technology, has given rise to sustainable agricultural supply chains. These supply chains enable farmers to utilize data properly and make valuable decisions [53].

However, sustainability is faced with some challenges: storage and logistics delivery, government regulations and policies [106]. Another challenge is in ensuring a sustainable supply chain system [91]. Establishing a reciprocal connection between logistics and production sustainability is also essential [119]. In the case of business 4.0 and logistics 4.0, it is imperative to carry out an environmental technical analysis of logistics network sustainable development [120].

### 4.1.3 Management

For performance management, the System Centre Operations Manager (SCOM) programs are developed to monitor an organization's software or hardware-based IT environment passively and actively manage information [121]. The present technological trends that lead to smart communications across all fields have improved management performance in terms of selling and buying using a smart contract for effective management distribution [74]. It plays a significant role in blockchain technology. Management uses these technologies to control the reverse auction process for improved performance [80]. Additionally, it is interesting to know that performance management helps to strengthen profits and downtime [103]. Complexity theory has been applied to strategic management and organizational studies [56]. This theory includes learning how organizations or companies respond to their circumstances and how they can deal with unpredictable situations, especially in conflicting situations [110].

The introduction of the IoT has dramatically enhanced management performance [122]. IoT has been an unprecedented scientific phenomenon for both scholarly and private organizations. IoT advertising is expanding every day because the IoT will offer infinite advantages to our surroundings.

The management process, including prioritized dangers for up-to-date decision-making, is linked to blockchain implementation [123]. Additionally, improper adaptabilities and compatibility features during the implementation phase of blockchain frameworks [124]. Moreover, identifying the fields best suited for implementing blockchain in their SCM [125] and the heterogeneity of the network environments also pose management challenges and different environments and standards for deployment [126]. Other challenges include assessing pressing questions on the adoption of the financial sector blockchain evaluation of risk factors in the organizational implementation of blockchain technologies, assessment of vehicular network stability of services for blockchain implementation and evaluation of trust and risk management for various suppliers of cloud services [127].

### *4.2 Improvements*

The increasing emphasis on new technology has raised several questions on the sustainable aspects. These improvements have enabled blockchain to lay the groundwork for a financial revolution in the oil, tourism, medicine, industry and supply chain [128]. Furthermore, blockchain implementation in the supply chain increases performance, reduces costs and strengthens connections between all stakeholders. It also builds more trust and optimizes the market processing involved. Examples of efficient blockchain deployment with the supply chain for wood and transport have been provided [129]. Blockchain has been applied in wood SC to provide traceability from cutting into functional materials. While blockchain was initially developed as Bitcoin infrastructure, several other applications in various industries, including finance, supply chains, IoT, authentication verification and data protection, for which blockchain technology has been used to improve fraud detection and effectiveness in many contexts [130].

### *4.2.1 Effectiveness*

Effectiveness is crucial in the blockchain supply chain, especially the ripple effect. The ripple effect is the idea that a single event affects several bodies. In order to help banks, rethink their cost structure, ripple created a cost model that aids in addressing current inefficiencies. Ripple also uses a trusted method to validate transactions through a group of servers instead of using blockchain mining. In contrast to Bitcoin transactions that use more power, take longer to validate, and contain high transaction costs, ripple transactions use less energy than Bitcoin, are verified in seconds and cost very little. For example, ripple effects in Big Data analysis increase promotional action consistency, improve consumer expectations, and raise awareness of the supply chain and promotions for consumer service. Likewise, in industry 4.0, like IoT, smart products and so on, ripple effects help to customize products with increased market stability, risk diversification, increased response time and improved power utilization, thereby leading to product integrity [95]. Data stored in a database can be protected through blockchain technologies. It can be done through the blockchain's well-formed transfers, verification, and auditing. The number of potential data integrity risks can be reduced [89]. While more effort is needed to better blockchain technology, the privacy, integrity and availabilities provided to our users are some of the positive elements of this technology. Integrity has been discussed in many articles because many organizations with centralized account privacy violations have caused a lack of customer trust and identity fraud [64].

Blockchain may be the remedy for the most stringent levels of data integrity. Blockchains are, by their nature, intrinsically resistant to data changes. Blockchain ledgers are permanent and cannot be changed or erased once data or transaction has been added. Moreover, blockchains are a data framework and a data system timekeeping tool, meaning data histories can accurately be reported and modified to the latter. Auditing organizations, regulatory enforcement standards and legal issues can use blockchain technologies to enhance and save money on data security [131].

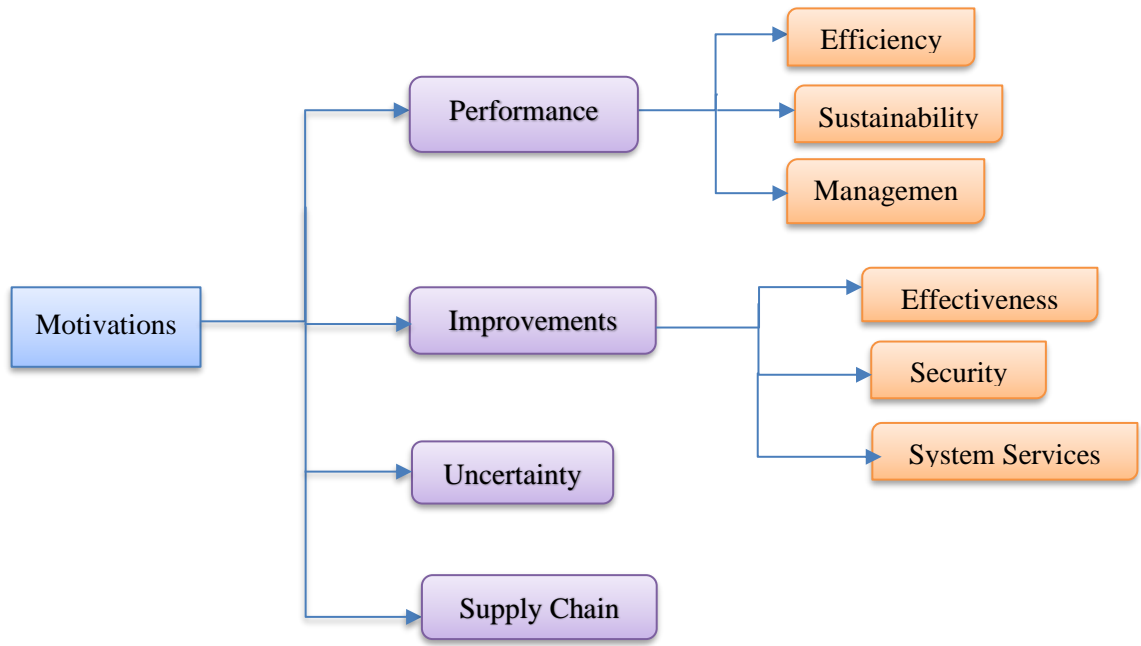

**Figure 3:** Categorization of Motivations

*4.2.2 Security*

Based on security, the Ethereum blockchain automated engine charging platform built on a network of multi-criteria decision support systems using the PROMETHEE approach was proposed by. This research aims to protect the flow of knowledge and money in the energy production-to-consumption phase. Energy manufacturers, customers, traders, retailers, electric charging stations and electric car users are all included in this report. Moreover, smart contracts are used to facilitate transactions between users. Smart contracts also remove the need for third-party contacts and all transactions involving the decentralized application are registered on the Bitcoin network [132]. Introduced a system for assessing the efficiency of an electricity blockchain using the fuzzy DEMATEL technique. In this research, the authors considered a distributed generation environment in which domestic consumers can generate power in their homes and purchase energy from the grid or other consumers. The blockchain allows the peer to acquire and sell and manage smart contracts.

Similarly, [133]proposed an innovative modelling approach for smart data discovery applications in the blockchain. The proposed model is also called 'A guidance approach. This article suggests a framework for evaluating new technologies and incorporates three scientific frameworks: the fundamental theory of complex information systems, systems theory and the International Organization for Standardization (ISO 25001) software quality standards.

With the availability of technology analytics, data analysis and statistical modelling, companies today have a greater understanding of creating more profitable and effective products. However, in the context of tourism and hospitality, [134] illustrates the critical role of big data in balancing operational goals with tourist needs by distinguishing and explaining the analytical mechanisms to support an integrated B2C interface based on numerous internal datasets and external data sources. The responsibility of stakeholders and the resources required are clarified and the full potential of big data in tourism and hospitality is demonstrated.

This research aims to investigate the ability of current digital technologies to increase healthcare quality and safety and to examine the evolving trend of digital medicine. Similarly, [88] proposes a novel framework to facilitate risk analysis in M-Health networks. The authors use a four-step approach, risk estimation, evaluation, archiving and risk parameter exchange. The accuracy of their results was checked and validated by comparing them to those of the Weighted Average process. The authors explored the efficacy of their method by adapting it to a specific Onion Attack. The eHealth framework has been suggested by [135] that deploys many instances of three-tier software patient agents, namely the sensing, the NEAR processing and the FAR processing layers, which enhance the confidence and defect-tolerant eHealth scheme.

Furthermore, a joint trust and risk model based on statistical analysis is presented [136]. The model, called JRTM, tackles not only security-related risks but also privacy and service efficiency risks. It is amenable to automated care, allowing for the representation of the service chain and the complex monitoring of risk levels based on profiles developed to distinguish between negative and positive performance based on cloud computing expectations in risk management.

Blockchain can also be used to solve the cyber protection issue in financial transactions as potential[137]. Analyses the ability of blockchain to improve the security of financial transactions using the Variable Frequency Transformer (VFT)-based objective framework. The VFT-based objective framework suggested in this paper enables an enterprise to expand this concept, beginning at the tactical level. An extensive survey of IoT attacks, security problems and blockchain solutions was provided by[138].

The authors highlight the essential blockchain-based technologies developed in recent times to address the issues faced by conventional cloud-centred applications[139]. Introduces a different Emission Trading Scheme (ETS) framework for Industry 4.0 integration. The proposed framework uses blockchain technologies and smart devices to boost ETS regulation policies. Blockchain characteristics, such as simplicity and immutability, can guarantee the data consistency used for the scheme. As a result, the scheme's functionality, continuity and integrity will be enhanced.

Compared to current related frameworks and checking against cyber injections in a real-world cyber-attack, the efficiency of the proposed system is demonstrated. Additionally, the study [140] addresses the primary issue of counterfeiting products in today's world. This study examines the capacity and demands of distributed ledger systems. SWARA (Step-wise Weight Assessment Ratio Analysis)-WASPS (Weighted Aggregate Amount Product Assessment) technique is used in the study[141]. SWARA is used to determine the weights of the parameters and WASPS is used to evaluate and prioritize the alternatives. Finally, the blockchain government information resource sharing and exchange model was proposed by [142], which consists of three parts: the network, infrastructure and business application layers. The proposed model is validated by five infrastructure networks to successfully address the problem of exchanging government knowledge services. It introduces novel approaches to topics such as trust islands in sharing, data ownership, peer management, standards compatibility and non-real-time exchange. This paper also develops a rigorous assessment model based on the TOPSIS framework and assesses the level of growth of smart Hefei from 2012 to 2017.

In policy-making, blockchain technology allows the company staff to communicate securely among several organizational units. For instance, the ability of blockchain to improve financial transaction protection was discussed by[143]. This is significant from a management perspective because businesses need to consider how blockchain technology may impact the online security of their financial transactions[144] and develop a system for examining challenges to blockchain acceptance and effective deployment in various industries and services. The authors describe these barriers using existing literature and expert views. Results from their studies can enable management to remove or address significant challenges. The proposed model increases the multiparty mechanism by showing that the decision-making components are of relative value in numbers obtained from Hierarchical Decision Models (HDMs). The knowledge-driven economy's tax, financial and social regulatory systems are essential. Blockchain algorithms for robust control and fog computing were proposed by[145]. It is a framework for evaluating the expense of government benefits for the general public. A consistent electoral system also focuses on direct or weighted voting. Blockchain technology also paves the way for anonymity and privacy across different domains.

*4.2.3 System Services*

Recent advancements in automobile technologies have resulted in a growth in their use. With the increased usage of electric cars, it is critical to protect the transfer of knowledge and money from manufacturing to consumption. While greenhouse gasses from petroleum-based oils pose global issues, Europe is facing a broad-based energy transformation replacing low-carbon, renewable energy sources such as wind and solar, fossil fuels and nuclear power. As a result, An Ethereum blockchain energy ecosystem is developed by [146] for recording all processes from electricity generation to end-users. Autonomous vehicles are becoming more common, and the features are already available in many consumer vehicles. AI becomes more prevalent in the real world with each mission. It is unavoidably a simulation component. The author of [147] proposed a framework for evaluating new technologies that employ a multi-criteria hierarchy. It focuses on understanding new technologies' evolving mechanisms and reducing errors.

Data entry, collection and sharing may be governed by specifying acceptable multilateral computer processable data sharing contracts. Thus, blockchain technology has also reduced errors drastically to their minimal form in smart objects and devices when performing operations. Several techniques that focus on error reduction in smart things are proposed. Among them are a Knowledge-Based Framework (KBS) based on smart devices and a data fusion paradigm to support industrial management decision-making in a clothing manufacturing enterprise proposed by [148]. The KBS helps to solve various types of decision problems, including factory tracking, preparation and monitoring of the manufacturing, efficiency management, real-time monitoring and acquisition and processing of data. Three real-world case studies of three software producers evaluated the decision model. The case study participants stated that the approach gives much greater insight into the selection process for blockchain platforms, provides a more prioritized list of options rather than having researched on their own and reduces the time and cost of decision-making. Moreover, the use of business process digitalization was discussed [149].

The authors introduced an approach that helps organizations to identify the digital technology best suited for a separate business process by combining Alternative Dispute Resolution (ADR) as an analysis model with SMEs as research methods. In this framework, shipowners and energy managers are exposed to barriers. It allows us to focus and solve crucial obstacles to improve energy quality consistently. Moreover, researchers and decision-makers can still use the system because it makes the energy conservation issues obvious.

*4.3 Uncertainty*

The innovation of blockchain technology has improved users' confidence and reduced doubts due to its robust security nature. Moreover, several researchers have acknowledged that blockchain has generally improved user trust in this technology. For example, the government blockchain resource sharing and exchange model proposed by [150] comprises three components: network, infrastructure and business implementation. Five service networks are funded to successfully resolve the exchange of knowledge services between governments. It offers new solutions in exchange for problems, such as confidence zones and continuity in expectations. A systematic analysis of fuzzy logic decisions was proposed by [151] to achieve a consistent ranking of alternatives: a new Pythagorean fuzzy linguistic multi-attribute decision-making model. Timely distribution measures the supplier's safety and requirements for developing resilient supply output profiles are explored.

The concept of the risk profile and resilient supply chain efficiency is theorized. They demonstrate that their methodology can effectively distinguish the correlations of deviations from the resilient supply chain value profile with the vulnerability performance profiles.
Furthermore, it presents [114] a new context-aware selection mechanism for Radio Access Technology (CRAT), which discusses device context and networks. In the NS3 modelling tool, the proposed CRAT was implemented and validated. The selection mechanism chooses the most suitable RAT to serve in an Ultra-Dense Network (UDN) environment.

### *4.4 Supply Chain*

Although several practical observations have been established in the given field recently, new research directions and ripple effect taxonomies have been identified[152]. The current network of supply chains consists of many phases (or stages of the supply chain), where several players compete for a market share. Each echelon is a monopoly, a duopoly, or an oligopoly in business. For instance, the blockchain domain, which comprises several players vying for its spectrum share, constitutes an oligopoly sector[153]. Moreover,[154] examines whether ratings of financial and vendors can be combined into a credit rating model of the supply chain. The authors further investigate how a paradigm of this kind is beneficial and challenging for all interested players. The author in [155] observation indicated a comprehensive approach to evaluating the supply chain and logistics sector's capacity for transition into more competitive economic environments while identifying new growth models to address possible macroeconomic stability. It is accomplished by an integrated evaluation of supply and demand processes that considers the core stakeholder interests. Unique perspectives for risk control disruption, such as the cascade impact and resilience of supply chains in business 4.0, have been introduced[120], emphasizing risk analysis for the supply chain.

Furthermore, [91]analyzes the use of BC technologies and an effort to improve effective, sustainable supply chain management (SSCM) instead of ineffective SCM. After conversations with academic and business experts, essential variables related to BC are defined from the literature. Those variables are further evaluated and formed using the Primary Component Analysis (PCA) laboratory for Fuzzy Decision (DEMATEL). The model of results for the integrated industrial strategy of 14.0 solutions for SMEs was developed by[156]. In this context, they consolidate the sixteen significant fields of operation and identify adequate output metrics to evaluate them[157]. This approach aims at creating a list of appropriate attributes for assessing the 14.0 solutions to be easier and more systemic. A multi-criteria optimization framework based on reluctant fuzzy is proposed by [158] to sets blockchain technologies in supply chain management. The author in [131] suggested a decision process for evaluating the feasibility of blockchain in logistics operations. This research focused mainly on the development employing a two-stage multi-criteria decision review of the viability of blockchain technology, particularly in logistics industries.

The integrated approach to using Blockchain technologies in the Indian agricultural supply chain was introduced by [120]. The work's uniqueness is three-phase: first, a reasonable number of individuals are involved who have helped finalize blockchain obstacles and enrich the literature with new information. The second element is the interconnected, mutually complementary approach to analysis. The third part of the work extends the proposed technique to large agricultural centres in developed countries, which require blockchain implementation to achieve food protection and safety. Furthermore, the TOPSIS multi-criteria decision model used to analyses risk safety and choose online transaction methods using Intuitionistic Fuzzy Shannon entropy weight was proposed by [104]. The proposed models are used to pick online payment methods based on many parameters relative to cryptocurrency bitcoin, the current online payment method. Similarly, [122] offered a blockchain service provider selection based on the best-worst way, the (BWM)-TOPSIS integrated process, in an intuitive environment. The proposed framework help enterprises estimate which blockchain vendor is more appropriate by considering more comprehensive influence factors.

## 5 Conclusion

This paper reviews previous studies that applied the smart contract analysis based on MCDM approaches to various challenges and motivations. This review paper aimed to classify challenges into four parts. The implementation performance is based on the system, security, and Integrity. The decision-making performance with the planning and execution steps. The challenges show the performance of a smart contract for the policy and applications.

Furthermore, the motivations act to improve the smart contract Implementation based on efficiency, sustainability, and Management. The improvements appear through analysis of the effectiveness, security, and system services. Finally, the uncertainty and supply chain performance improved users' confidence in offering new solutions in exchange for problems in the smart contacts.

**Funding Statement:** No external entity is supporting this investigation.

**Conflicts of Interest:** The authors declare that they have no conflicts of interest to report regarding the present study.